\documentclass[12pt]{article}
\usepackage[dvips]{graphicx}
\usepackage{color}
\usepackage{amsfonts,amssymb}
\begin{document}

\vspace*{1cm}
\begin{center}
{\Large \bf On the BELLE Charmonium States}\\

\vspace{4mm}

{\large A. A. Arkhipov\footnote{e-mail: arkhipov@mx.ihep.su}\\
{\it State Research Center ``Institute for High Energy Physics" \\
 142281 Protvino, Moscow Region, Russia}}\\
\end{center}

\vspace{2mm}
\begin{abstract}
It is shown that newly performed experimental studies by the Belle
Collaboration are excellently incorporated in the unified picture for
hadron spectra developed early. From our analysis it follows that the
measured values for the masses of the BELLE states exactly coincide
with the calculated masses of the states living in the corresponding
KK towers.
\end{abstract}

\section*{}

Quite recently the Belle Collaboration \cite{1} reported the
observation of an enhancement in the $\omega J/\psi$ invariant mass
distribution for exclusive $B\rightarrow K \omega J/\psi$ decays. The
statistical significance of the $\omega J/\psi$ mass enhancement was
estimated to be greater than $8\sigma$. The results were obtained
from a 253~fb$^{-1}$ data sample that contains 274 million $B\bar{B}$
pairs that were collected near the $\Upsilon(4S)$ resonance with the
Belle detector at the KEKB asymmetric energy $e^+e^-$ collider.

The reported fits to the individual 40 MeV-wide bins of $M(\omega
J/\psi)$ are plotted in Figs.~1(a) and (b) extracted from original
paper \cite{1}. An enhancement is evident around $M(\omega
J/\psi)=3940$~MeV. The curve in Fig.~1(a) is the result of a fit with
the same threshold function which has been used to fit the
phase-space MC distribution. The fit quality  in that case is poor
($\chi^2/dof = 133/11$), indicating a significant deviation from
phase-space.

Figure~1(b) shows the results of a fit where an $S$-wave Breit Wigner
function has been included to represent the enhancement; see,
however, the details in  \cite{1}. The fit with $\chi^2/dof = 18.8/8$
(CL = 1.6\%) yielded a Breit-Wigner signal with a mass $M = 3941\pm
11$~MeV and a width $\Gamma  = 92\pm 24$~MeV (statistical errors
only). It was pointed out that the Breit Wigner fit parameters are
sensitive to the shape used to model the background. For example, if
the background function is replaced with a second-order polynomial
function, then the mass is changed to $3949\pm 9$~MeV, the width is
changed to $\Gamma = 112\pm 25$~MeV and the fit quality is improved:
$\chi^2/d.o.f. = 11.8/6$ (CL=6.7\%). For the average from the two
fits one obtains a mass $M = 3945\pm 10$~MeV and a width $\Gamma  =
102\pm 24.5$~MeV.

Earlier, in September of 2003 at the 10th International Conference on
Hadron Spectroscopy (Ashaffenburg, Germany, 31 August -- 6 September
2003), the Belle Collaboration reported \cite{2,3,4} the discovery of
very narrow $X(3872)$-meson state with a mass of
3872.0$\pm$0.6(stat)$\pm$0.5(syst) MeV ($\Gamma_{X(3872)}^{tot}<2.3
MeV$ 95\% C.L.) in the $\pi^+\pi^-J/\psi$ invariant mass distribution
in the $B$ decay $B^{\pm} \rightarrow K^{\pm}\pi^+\pi^-J/\psi$, where
it was also established that the $\pi^+\pi^-$ invariant mass for the
$X(3872)$ signal region concentrated at the $\rho$ mass. This
observation of the Belle Collaboration was soon confirmed by CDF at
Fermilab \cite{5}. The mass measurement presented by CDF 3871.4 $\pm$
0.7 $\pm$ 0.4 MeV is in agreement with the result of Belle. It should
be noted, in particular, that the mass 3872 MeV is very near the
$D^0{\bar D}^{*0}$ threshold, while the $D^+D^{*-}$ channel with
approximately 8 MeV higher threshold mass is forbidden for $X(3872)$
decay by phase space. What is non-trivial  here that the $X(3872)$
state has been observed at the mass which is surprisingly far from
the predictions of conventional quark potential models. It is still
more remarkable that the observed state $X(3872)$ was very narrow.
The small width was found to be in contradiction with quark model
expectations too.

Many attempts have been made to understand the $X(3872)$ state as a
simple $c\bar c$ charmonium state. A comparative analysis of many
possible $c\bar c$ assignments for the $X(3872)$ state has been
presented in comprehensive paper \cite{6} where some non-$c\bar c$
$X(3872)$ assignments have been considered as well. However, the
mechanism which keeps this state narrow is unclear so far. It is even
quite unclear how the $X(3872)$ meson could be understood as the
state of a simple $c\bar c$ quark system, this is extremely
problematic in the framework of phenomenology based on conventional
quark models. The most quark potential models predict 1D states about
50--100 MeV below the $X(3872)$ mass, and the 2P states are predicted
to lie above the $X(3872)$ by a similar amount \cite{6}. That is to
say, the $X(3872)$ mass is a subject that is to be explained in the
conventional quark potential models.

In our talk \cite{7} presented at the same 10th International
Conference on Hadron Spectroscopy, where some of our previous studies
were partially summarized, we have reported a new theoretical concept
to create quite a new scheme of systematics for hadron states
providing a unified picture for hadron spectra. The fundamental
Kaluza-Klein hypothesis on existence of the extra dimensions with a
compact internal extra space together with some novel dynamical ideas
have been taken as a base of our approach to hadron spectroscopy. In
our theoretically developed concept the observed hadron states occupy
the storeys and live in the corresponding KK towers built in
according to the established general physical law. We have pointed
out that the BELLE $X(3872)$ state was excellently incorporated in
the systematics provided by the created unified picture for hadron
spectra. Such state really exists, and it lives just on the second
storey in the Kaluza-Klein tower of KK excitations for the $\rho
J/\psi$-system; see Table 1. As is seen from Table 1, there is a
wonderful agreement of experimentally measured mass with
theoretically calculated one.

In this note we present the build of the Kaluza-Klein tower of KK
excitations for the $\omega J/\psi$-system. In according to the
general physical law we built the Kaluza-Klein tower of
KK-excitations for the $\omega J/\psi$ system by the formula

\begin{equation}\label{omegaJpsi}
M_n^{\omega J/\psi} = \sqrt{m_{\omega}^2+\frac{n^2}{R^2}} +
\sqrt{m_{J/\psi}^2+\frac{n^2}{R^2}}\,,\quad (n=1,2,3,...),
\end{equation}
where $R$ is the same fundamental scale established before; see
\cite{7} and references therein for the details, and $m_{\omega}$ =
782.57 MeV, $m_{J/\psi}$ = 3096.87 MeV have been taken from PDG. The
such built Kaluza-Klein tower is shown in Table 2. As seen in Table
2, the Belle measured state just lives in 7th storey within this
Kaluza-Klein tower.

Our conservative estimate for the widths of KK excitations looks like
\begin{equation}\label{width}
\Gamma_n \sim \frac{\alpha}{2}\cdot\frac{n}{R}\sim 0.4\cdot n\,
\mbox{MeV},
\end{equation}
where $n$ is the number of KK excitation, and $\alpha \sim 0.02$,
$R^{-1}=41.48\,\mbox{MeV}$ are known from our previous studies
\cite{7}. This gives $\Gamma_{2}(X(3782)\rightarrow\rho J/\psi)\sim
0.8$ MeV and $\Gamma_{7}(X(3945)\rightarrow\omega J/\psi)\sim 2.8$
MeV. The Belle measured width $\Gamma_{exp}(X(3945)\rightarrow\omega
J/\psi)\sim 100$ MeV is too broad and cannot simply be explained. In
general, the broad peaks in the hadron spectra are interpreted as
envelope of the narrow peaks predicted in our approach, or some
additional model-dependent assumptions are needed. As it follows from
Table 2, we predict the narrow states $M_6^{\omega J/\psi}$(3928) and
$M_8^{\omega J/\psi}$(3965) which are near the observed state
$M_7^{\omega J/\psi}$(3945). However, the experimentally used 40
MeV-wide bins of $M(\omega J/\psi)$ cannot obviously resolve these
predicted narrow states. This only means that a further, much more
careful experimental studies with a higher statistics and better mass
resolution are very desired.

In summary, newly performed experimental studies by the Belle
Collaboration are excellently incorporated in the unified picture for
hadron spectra developed early. We would like to emphasize once again
that the main advantage of our approach to hadron spectroscopy is
that all calculated numbers for the masses of hadron states do not
depend on a special dynamical model but follow from fundamental
hypothesis on existence of the extra dimensions with a compact
internal extra space. Our analysis shows that the measured values for
the masses of the BELLE states exactly coincide with the calculated
masses of the states living in the corresponding KK towers. We expect
that new experiments with better statistics and higher mass
resolution will appear in the near future to extend our knowledge and
to enrich our understanding the true nature and the properties of the
newly discovered states.

\newpage
\vspace*{2cm}

\begin{figure}[htb]
\begin{center}
\includegraphics[width=\textwidth]{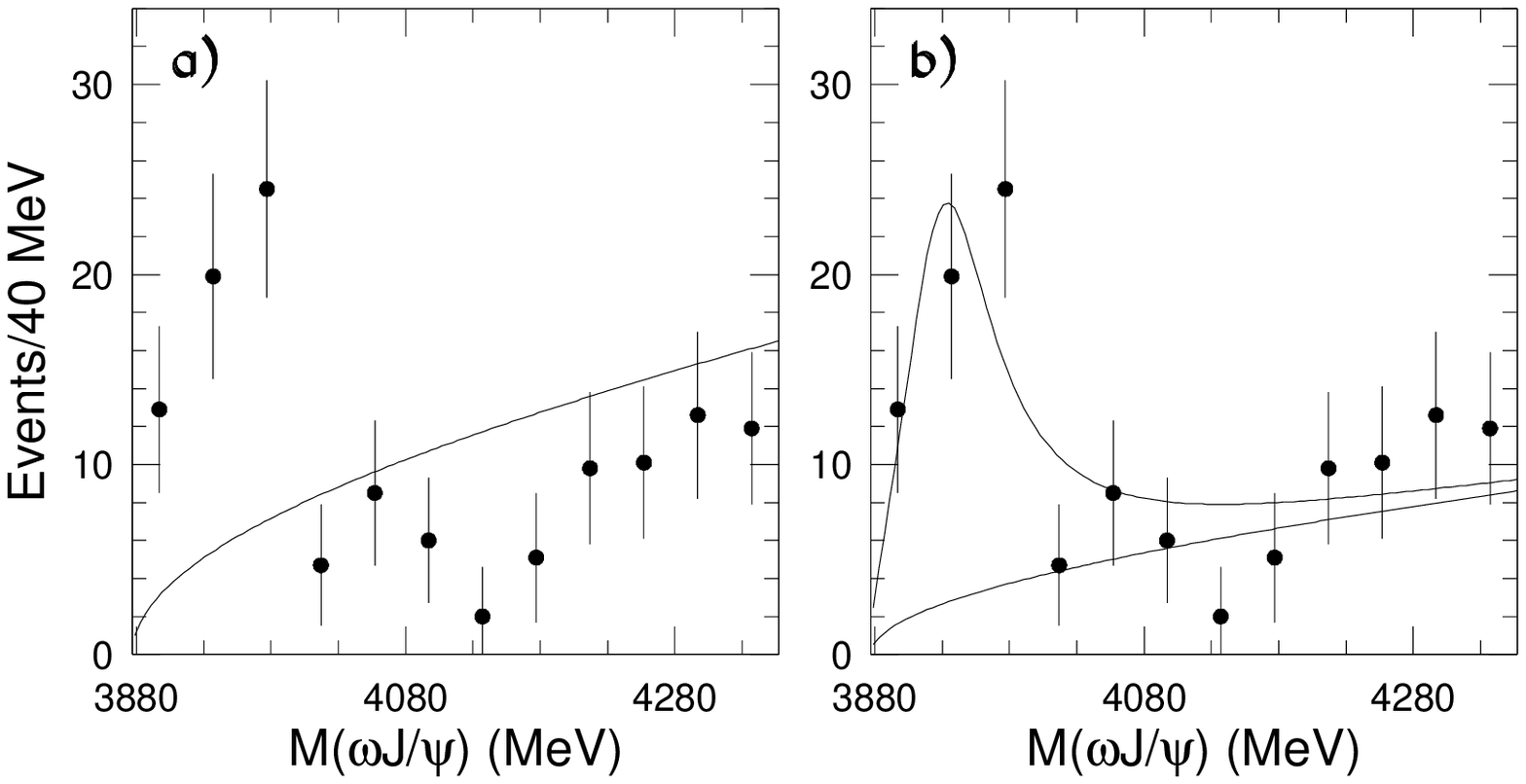}\label{fig1}
\end{center}
\caption{The $\omega J/\psi$ invariant mass in $B\rightarrow K\omega
J/\psi$ decay presented in Ref. \cite{1}. The curve in {\bf (a)}
indicates the result of a fit that includes only a phase-space-like
threshold function. The curve in {\bf  (b)} shows the result of a fit
that includes an $S$-wave Breit-Wigner resonance term. The details
for the fits shown see in original paper \cite{1}.}
\end{figure}

\newpage
\vspace*{1cm}
\begin{center}
Table 1. Kaluza-Klein tower of KK excitations for $\rho J/\psi$
system\\ and $D_{sJ}$(3872)-meson.\footnote{The Table extracted from
Ref. \cite{7}. The recent value for the Belle measurements of the
X(3872) mass is 3872.0$\pm$0.6(stat)$\pm$0.5(syst) MeV.}

\vspace{5mm}
\begin{tabular}{|c|c|c|c|}\hline
 n & $ M_n^{\rho J/\psi}$ MeV & $M_{exp}^{\rho J/\psi}$ MeV  \\
 \hline
1  & 3867.57 &   \\
2  & 3871.74 & 3871.8$\pm$0.7$\pm$0.4  \\
3  & 3878.67 &   \\
4  & 3888.30 &   \\
5  & 3900.58 &   \\
6  & 3915.41 &   \\
7  & 3932.73 &   \\
8  & 3952.42 &   \\
9  & 3974.39 &   \\
10 & 3998.54 &   \\
11 & 4024.75 &   \\
12 & 4052.92 &   \\
13 & 4082.95 &   \\
14 & 4114.74 &   \\
15 & 4148.19 &   \\
16 & 4183.22 &   \\
17 & 4219.74 &   \\
18 & 4257.68 &   \\
19 & 4296.95 &   \\
20 & 4337.48 &   \\
21 & 4379.23 &   \\
22 & 4422.11 &   \\
23 & 4466.09 &   \\
24 & 4511.11 &   \\
25 & 4557.11 &   \\
26 & 4604.07 &   \\
27 & 4651.93 &   \\
28 & 4700.65 &   \\
29 & 4750.21 &   \\
30 & 4800.57 &   \\ \hline
\end{tabular}
\end{center}

\newpage
\vspace*{2cm}
\begin{center}
Table 2. Kaluza-Klein tower of KK excitations for $\omega J/\psi$
system.

\vspace{5mm}
\begin{tabular}{|c|c|c|c|}\hline
 n & $ M_n^{\omega J/\psi}$ MeV & $M_{exp}^{\omega J/\psi}$ MeV  \\
 \hline
1  & 3880.82 &   \\
2  & 3884.94 &   \\
3  & 3891.77 &   \\
4  & 3901.28 &   \\
5  & 3913.40 &   \\
6  & 3928.05 &   \\
7  & 3945.16 & 3945 $\pm$ 10  \\
8  & 3964.62 &   \\
9  & 3986.35 &   \\
10 & 4010.24 &   \\
11 & 4036.18 &   \\
12 & 4064.09 &   \\
13 & 4093.84 &   \\
14 & 4125.36 &   \\
15 & 4158.54 &   \\
16 & 4193.31 &   \\
17 & 4229.56 &   \\
18 & 4267.24 &   \\
19 & 4306.26 &   \\
20 & 4346.55 &   \\
21 & 4388.05 &   \\
22 & 4430.71 &   \\
23 & 4474.46 &   \\
24 & 4519.26 &   \\
25 & 4565.06 &   \\
26 & 4611.82 &   \\
27 & 4659.48 &   \\
28 & 4708.02 &   \\
29 & 4757.41 &   \\
30 & 4807.60 &   \\ \hline
\end{tabular}
\end{center}

\end{document}